\documentclass[preprint,aps,showpacs]{revtex4}
\usepackage{graphicx}
\usepackage{tabularx}

\newcommand{\ba}{\begin{eqnarray}}
\newcommand{\ea}{\end{eqnarray}}
\begin{document}

\title{Evidence for Triangular ${\cal D}_{3h}$ Symmetry in $^{12}$C}

\author{D.J. Mar{\'{\i}}n-L\'ambarri$^1$, R. Bijker$^2$, M. Freer$^1$, M. Gai$^{3,4}$, Tz. Kokalova$^1$, D.J. Parker$^1$, C. Wheldon$^1$}
\affiliation{1. School of Physics and Astronomy~,~University of Birmingham~,~Birmingham B15 2TT, UK \\
2. Instituto de Ciencias Nucleares, Universidad Nacional Aut\'onoma de M\'exico, A.P. 70-543, 04510 M\'exico, D.F., Mexico\\
3. LNS at Avery Point,University of Connecticut~,~Groton,~CT 06340-6097, USA \\
4. Wright Lab, Dept of Physics, Yale University, New Haven, CT 06520-8124, USA}

\begin{abstract}

We report a measurement of a new high spin J$^\pi$ = 5$^-$ state at 22.4(0.2) MeV in $^{12}$C
which fits very well to the predicted (ground state) rotational band of an oblate equilateral
triangular spinning top with a ${\cal D}_{3h}$ symmetry characterized by the sequence
$0^+$, $2^+$, $3^-$, $4^\pm$, $5^-$ with almost degenerate $4^+$ and $4^-$ (parity doublet) states.
Such a ${\cal D}_{3h}$ symmetry was observed in triatomic molecules and it is observed here for the first time in nuclear physics.
We discuss a classification of other rotation-vibration bands in $^{12}$C such as the ($0^+$) Hoyle band
and the ($1^-$) bending mode band and suggest measurements in search of the predicted (``missing'') states
that may shed new light on clustering in $^{12}$C and light nuclei. In particular the observation (or non-observation) of the predicted (``missing") states in the Hoyle band will allow us to conclude the geometrical arrangement of the three alpha-particle composing the Hoyle state at 7.654 MeV in $^{12}$C.

\end{abstract}


\maketitle

Geometrical equilateral triangular configurations \cite{Bi00,Bi02} have been identified in the triatomic H$_3^+$ molecule \cite{X3} where the predicted spectrum of a triangular oblate spinning top with a ${\cal D}_{3h}$ Symmetry was observed \cite{Bi00,Bi02,X3}.
It was suggested \cite{Bi00,Bi02} that the three alpha-particle system of $^{12}$C should lead
to similar ``triatomic like" structure in nuclei. The application to $^{12}$C of the ${\cal D}_{3h}$ Symmetry, a mathematical tool that was developed to describe molecular structure, emphasizes the role of symmetry across very different energy scales, and it leads to a model of $^{12}$C that correctly predicts several new observations that we report for the first time in this letter. Such a poly-atomic-like description of light nuclei should lead to a better understanding of the clustering phenomena in light nuclei. 

In this letter we demonstrate that this U(7) model \cite{Bi00,Bi02} as applied to $^{12}$C predicts all known low spin (cluster) states below 15 MeV, as well as the measured B(E$\lambda$) and form factors measured in electron scattering \cite{Tum00}. But perhaps more importantly the model predicts new (``missing") states, the observation (or non-observation) of which will allow us to resolve a major problem of current concern on the geometrical arrangement of the three alpha-particles in the Hoyle state at 7.654 MeV. We demonstrate that the observed rotation-vibration spectrum of $^{12}$C in of itself already indicates the geometrical structure of $^{12}$C.

We report the observation of a new $J^\pi = 5^-$ state that fits very well the
rotational $J(J+1)$ trajectory of the ground state band of $^{12}$C as predicted by the U(7) model. In addition, the 
$4^-$ state recently observed by some of us at 13.35 MeV in $^{12}$C \cite{Fre07,Kir10} confirms the $J^\pi = 4^\pm$ parity doublet predicted by this U(7) model for the ground state band of $^{12}$C. The ground state rotational band including the states of J$^\pi$ = $0^+, \ 2^+, \ 3^-, \ 4^\pm$ and $5^-$ is a strong signature of a ${\cal D}_{3h}$  symmetry and it is observed here for the first time in a nucleus. 

The triatomic U(7) mixed-parity structure observed in $^{12}$C resembles the diatomic U(4) mixed parity structure \cite{U4}
observed in $^{18}$O \cite{18O} and the Tetrahedral Symmetry ($T_d$) mixed parity structure recently observed in $^{16}$O \cite{16O}. 

The structure of $^{12}$C has recently attracted much theoretical attention due to the availability
of {\it ab initio} no-core shell model calculations \cite{abinitio}, the no-core symplectic model
\cite{Draayer} and Effective Field Theory (EFT) calculations on the lattice \cite{EFT}. These calculations
attempt to provide a microscopic description of cluster states that are well described in the traditional
clustering model \cite{Kamim} and Antisymmetrized Molecular Dynamics (AMD) \cite{AMD}, as well as in the more modern Fermionic Molecular Dynamics (FMD) model \cite{FMD} and more exotic cluster models \cite{BEC}. However, thus far {\it ab initio} shell model calculations
have failed to predict \cite{abinitio} the Hoyle state at 7.654 MeV in $^{12}$C that is known to be one
of the best examples of alpha-clustering in nuclei. 

The EFT lattice calculations \cite{EFT} and the FMD model \cite{FMD} predict an equilateral arrangement of the three alpha-particles in the ground state of $^{12}$C and hence they provide the microscopic foundation of the conjectured ${\cal D}_{3h}$ Symmetry of the ground state of $^{12}$C. But these models are currently unable to predict the high spin $5^-$ state reported here or the $4^\pm$ parity doublet that we observe in $^{12}$C.

The identification of the rotational excitations of the Hoyle state with $2^+$ \cite{2+} and $4^+$ \cite{Fre11} raises
an intriguing question of current concern regarding the geometrical structure of the Hoyle state, whether it is a bent-arm
configuration \cite{EFT} or rather an equilateral triangular configuration just as the ground-state.
In this Letter we point out that future measurements of predicted (``missing") rotation-vibration states in $^{12}$C will allow us to understand the geometrical arrangement of the three alpha-particles in the Hoyle state-- a problem which is as old as the discovery of the Hoyle state itself.

The present measurements were performed at the Birmingham MC40 cyclotron facility. A beam of $^{4}$He
nuclei at an energy of 40 MeV was incident on a $100~ \mu g/cm^{2}$ carbon target.
The reaction of interest was $^{12}$C$(^{4}$He,$3\alpha)^{4}$He,
in which the $^{12}$C nuclei were excited above the $\alpha$-decay threshold through the inelastic scattering process. 
An array of four, $500$ $\mu m$
thick, silicon strip detectors was used in order to detect three of the four final state
$\alpha$-particles. Each detector had a surface of $5 \times 5$ cm$^{2}$ subdivided into $16$ horizontal
and $16$ vertical strips, front and back, respectively. The detectors were placed at distances $13.0$,
$11.0$, $11.0$, $13.0$ cm from the target at angles $62.0^{\circ}$, $32.0^{\circ}$, $-32.0^{\circ}$,
$-62.0^{\circ}$, respectively (the signs indicate opposing sides of the beam axis).
The array covered an angular range of $\theta_{lab}=$ $20^{\circ}$ to $75^{\circ}$.
The detection system was calibrated with a triple $\alpha$-particle source. These detectors allowed the energy and emission angle of each particle 
to be determined and hence the momentum, assuming each to be an $\alpha$-particle.

\begin{figure}
   \begin{center}
    \includegraphics[width=1.0\textwidth]{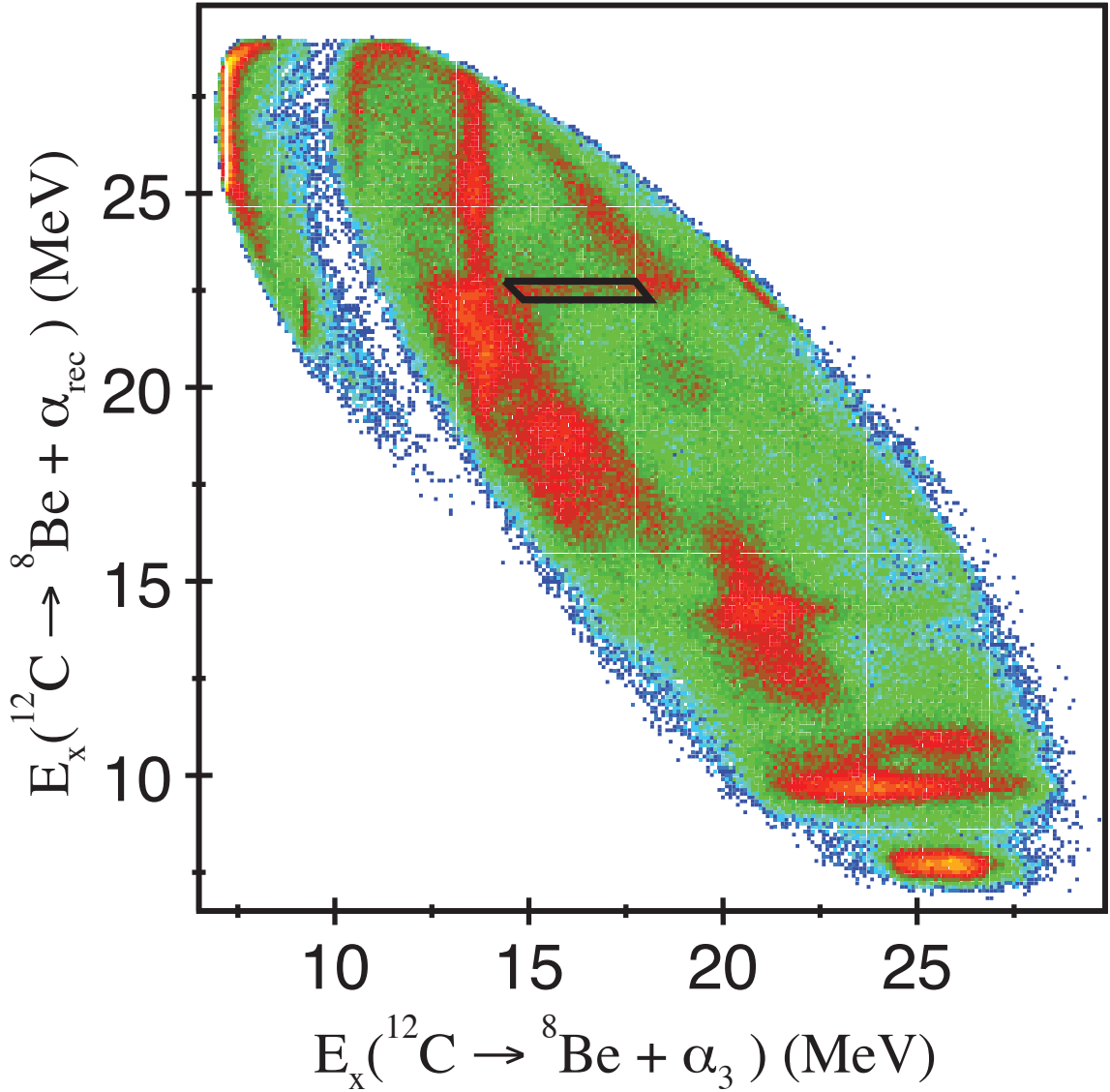}
     \end{center}
\caption{(Color online) Dalitz plot for the $^{12}$C($^{4}$He,$^{12}$C*)$^{4}$He reaction. Excitation energies of $^{12}$C are plotted on the horizontal axis in which three $\alpha$-particles were detected. On the vertical axis the $^{12}$C excitation energies are calculated by the reconstruction of the undetected $\alpha$ particle and the $^{8}$Be nuclei. The tilted square box shows the region selected for the angular distribution analysis.} \label{figexp1}
\vspace*{-0.5cm}
\end{figure}

The 4${th}$ undetected $\alpha$-particle's properties
were reconstructed using conservation of energy and momentum. 
Events in which any two of the three detected $\alpha$-particles resulted from the decay of $^{8}$Be$_{gs}$
were selected. In order to reconstruct the origins of the final state particles a Dalitz plot was created as shown in Fig.~\ref{figexp1}. Here the
excitation energy in $^{12}$C was calculated by reconstructing $E_x(^{12}$C) from the $^8$Be and either the detected (horizontal axis) or undetected (vertical axis) $\alpha$-particle. The horizontal and vertical loci correspond to $^{12}$C excited states and the weak diagonal loci states in $^8$Be. 

\begin{figure}
   \begin{center}
    \includegraphics[width=1.0\textwidth]{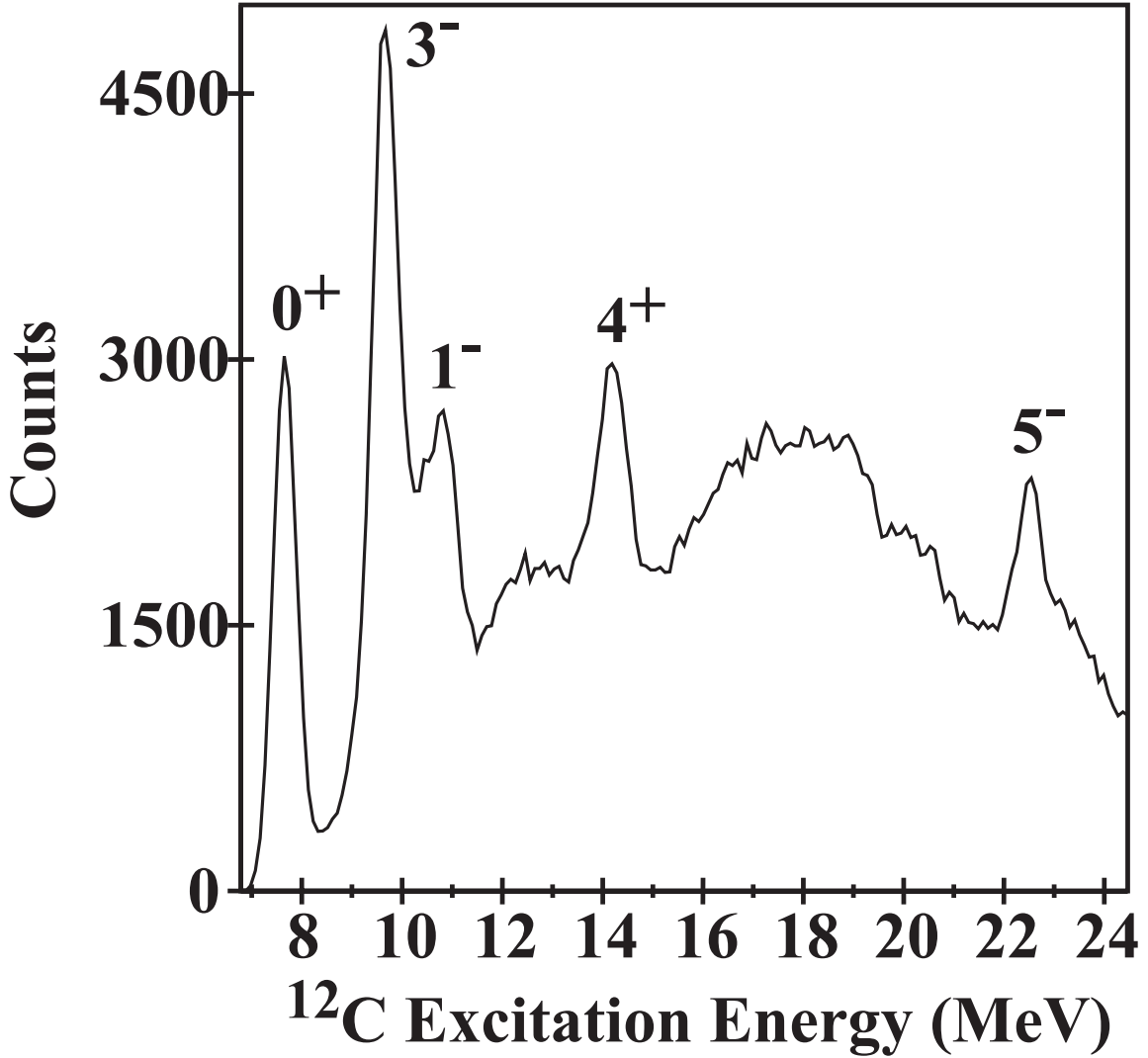}
     \end{center}
\caption{Projection of the Dalitz plot in Fig.~\ref{figexp1} onto the vertical axis. In addition to the known states a peak is 
observed at 22.4(0.2) MeV. The broad background (between 12 and 20 MeV) is due to "leaking" of excited states of $^8$Be into the projected region.} \label{figexp2}
\vspace*{-0.5cm}
\end{figure}

The projection of the Dalitz plot onto the vertical axis is shown in Fig.~\ref{figexp2}. States at 7.654 (0$^+$), 9.641 (3$^-$), 10.844 (1$^-$), 14.083 MeV (4$^+$) are observed. In Figs.~\ref{figexp1} and~\ref{figexp2} it is also
possible to observe a peak which would correspond to a state at $22.4(0.2)$ MeV which has not been previously reported.

\begin{figure}
   \begin{center}
    \hspace*{-0.5cm}
    \includegraphics[width=1.0\textwidth]{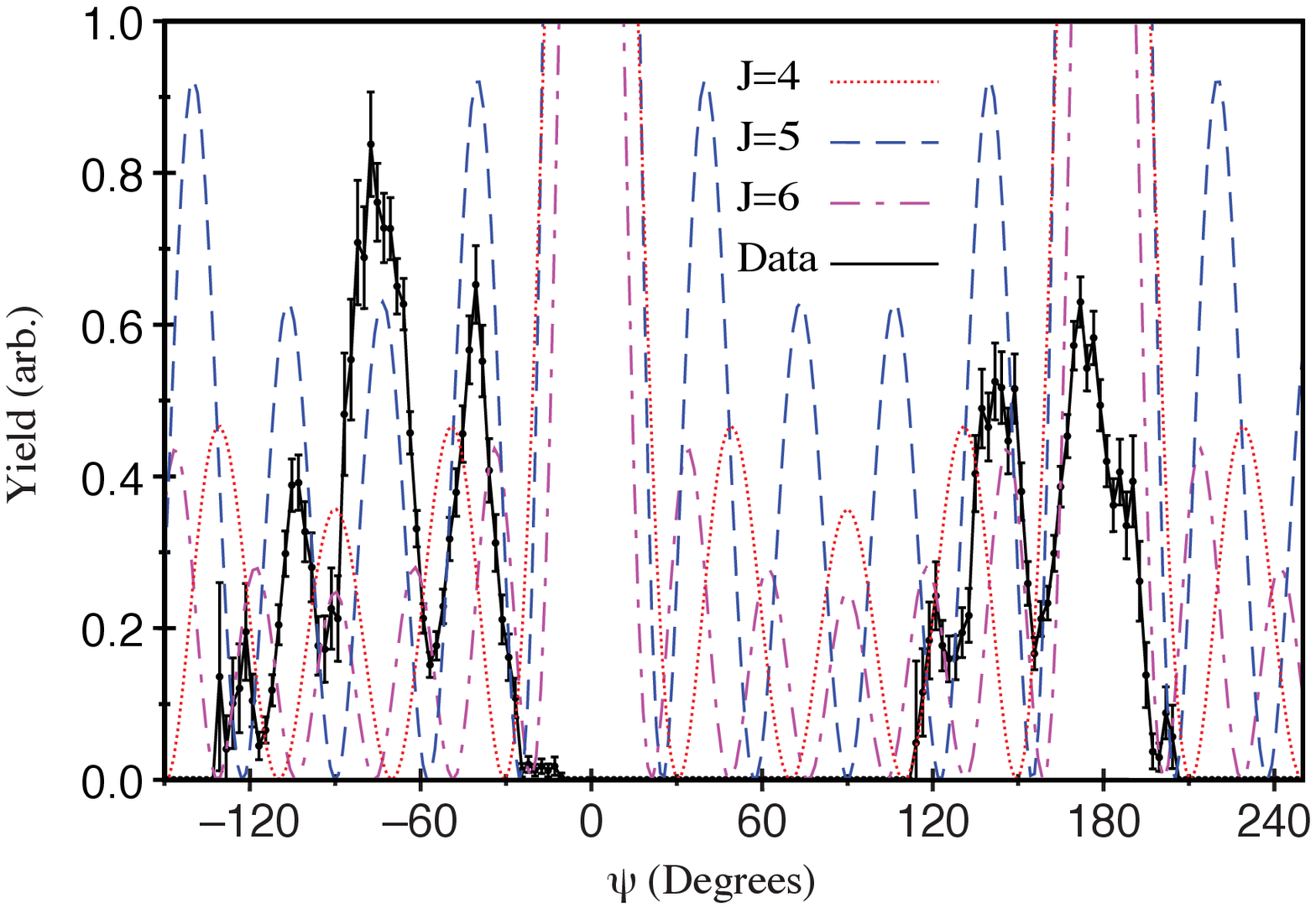}
     \end{center}
\caption{(Color online) The projection onto the $\psi$ axis of the angular correlations for the $22.4$ MeV state. The data points are corrected for the acceptance of the detectors and connected with a (continuous black) line to guide the eye. They are compared with the Legendre polynomials $|$P$_{5}$[cos($\psi)]|^{2}$ (dashed blue line) as well as for $\ell=4$ (dotted red line) and $\ell=6$ (dotted-dashed red line). Note that due to the unknown m-substate population of the J$^\pi = 5^-$ state the height of the oscillations cannot be predicted, but the oscillatory phase determines the angular correlation to arise from a J$^\pi = 5^-$ state.} \label{figexp3}
\vspace*{-0.5cm}
\end{figure}

In order to achieve a characterisation into the nature of the proposed new state in $^{12}$C we have
used the angular correlations technique; a method in which the distribution pattern of the products are analyzed \cite{Fre96}. This method yields a model independent spin determination when the initial and final state particles are all spin zero, and is described in more detail in
\cite{Fre11,Fre96}. Using the Dalitz plot in Fig.~\ref{figexp1}, it is possible to set a window around the specific data
of interest (shown by the tilted square box) and then generate the angular correlation plot for the selected events (similar to those in Ref.~\cite{Fre11}) for the proposed 22.4 MeV state. 

In the $^{12}$C($^{4}$He,$^{12}$C*)$^{4}$He reaction there are two center-of-mass frames, the first corresponding to the inelastic excitation, the second to the decay of $^{12}$C into $^8$Be+$\alpha$. The emission angle of the $^{12}$C decay process with respect to the beam-axis is described by the angle $\psi$ which is explicitly in the center of mass of the $\alpha + ^8Be$ system, hence arising from a state in $^{12}$C and not in $^{16}$O . For a $^{12}$C state of spin $J$, it would be expected that the angular correlation should oscillate with a period given by $|$P$_{J}$[cos($\psi)]|^{2}$. As described in Ref.~\cite{Fre11,Fre96}, it is possible to infer from the oscillation pattern of the data the spin of the excited state. The dependence of the yield on the angle $\psi$ is shown in Fig.~\ref{figexp3}, in which the data are compared with several Legendre polynomials. The measured alpha-spectrum and angular correlation clearly point to the existence of a state at 22.4(0.2) MeV with $J^\pi$=5$^-$.

\begin{figure}
\includegraphics[width=6in]{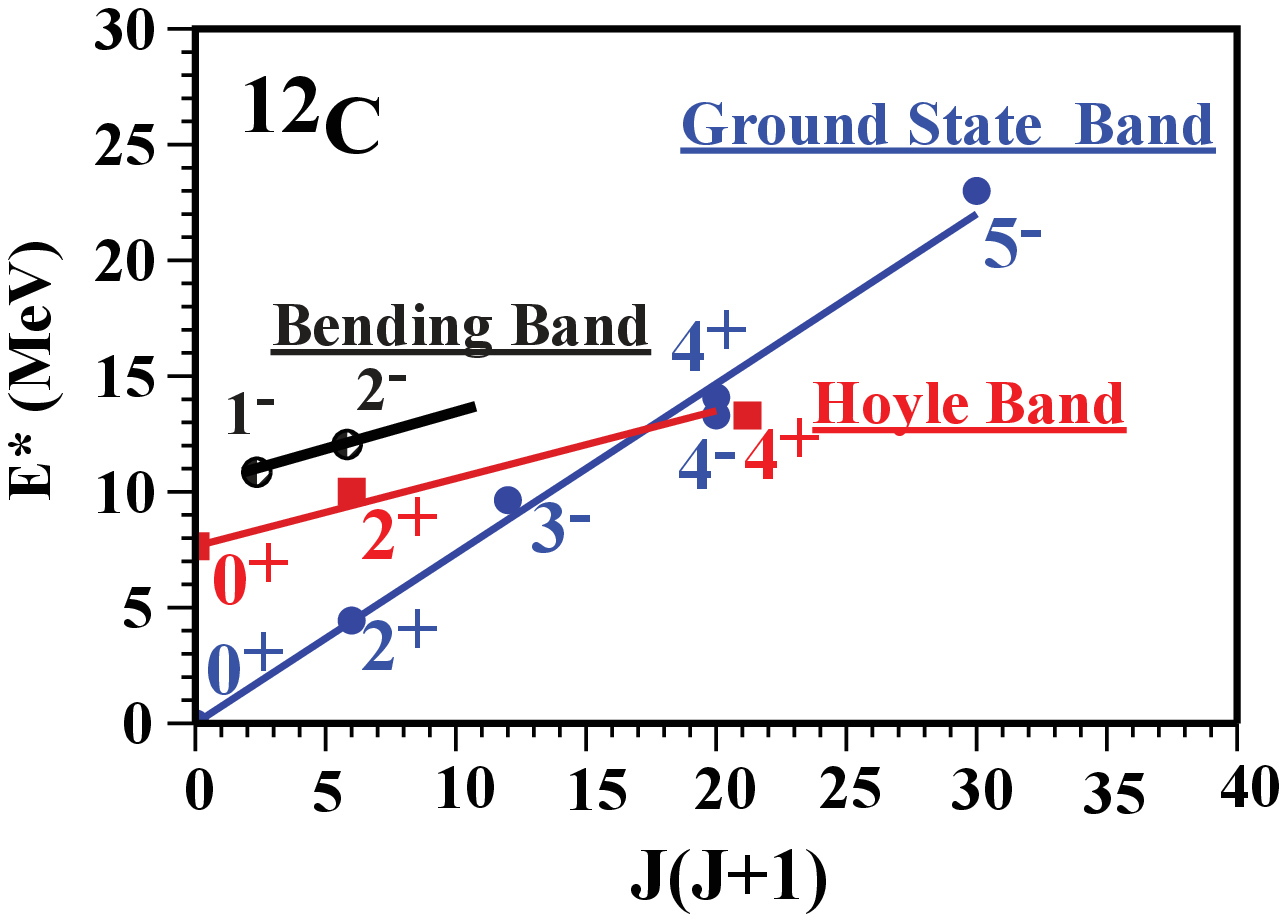}
\caption{\label{RotBands} (Color Online) Rotational band structure of the ground-state band, the Hoyle band and
the bending vibration in $^{12}$C.}
\end{figure}

In Fig.~\ref{RotBands} we show the rotational band structure in $^{12}$C. The ground state rotational band consisting
of the levels $0^+$, $2^+$, $3^-$, $4^\pm$ and the newly measured $5^-$ state, follow a $J(J+1)$ trajectory. Also the
recently identified rotational excitations with $2^+$ \cite{2+} and $4^+$ \cite{Fre11} of the Hoyle state form a $J(J+1)$ sequence albeit with
a larger moment of inertia. Finally as we discuss below, the negative parity states $1^-$ and $2^-$ shown in Fig.~\ref{RotBands} are assigned as members of the bending vibration with almost the same moment of inertia as the Hoyle band.

We present an analysis of the cluster states in $^{12}$C in terms of
oblate symmetric top which is a special case of the algebraic cluster model \cite{Bi00,Bi02}.
In this approach, the three alpha-particles are located at the corners of an equilateral triangle. Their relative
motion is described by two perpendicular Jacobi vectors, $\vec{\rho}$ and $\vec{\lambda}$, one vector connecting
two points on the triangle and the second one along the half angle perpendicular to it. The corresponding algebraic
model describing such a system is based on the $U(6+1) = U(7)$ spectrum-generating algebra \cite{Bi00,Bi02}.

Of particular interest is the oblate symmetric top limit which corresponds to the geometric
configuration of three $\alpha$-particles located at the vertices of an equilateral triangle.
The rotation-vibration wave functions of a triangular configuration can be written as \cite{Bi00,Bi02}
\begin{equation}
\mid N, (v_{1},v_{2}^{\ell_{2}}),K,L^{P} \rangle ~.
\end{equation}
Here N is the total number of bosons. The energy spectrum consists of a series of rotational bands labeled by
$(v_1,v_2^{\ell_2})$. Here $v_1$ corresponds to the breathing vibration with $A$
symmetry and $v_2$ to the doubly degenerate bending vibration with $E$ symmetry;
$\ell_2$ denotes the vibrational angular momentum of the doubly degenerate vibration,
$L$ the angular momentum, $K$ its projection on the symmetry axis and $P$ the parity.
Since we do not consider the excitation of the $\alpha$-particles, the wave functions
describing the relative motion have to be symmetric, {\it i.e.} $|K \mp 2\ell_2|=3m$
a multiple of 3 \cite{Bi00,Bi02}.
This imposes some conditions on the allowed values of the angular momenta and parity.
For vibrational bands with $(v_1,0^{0})$, the allowed values of the angular momenta
and parity are $L^P=0^+$, $2^+$, $4^+$, $\ldots$, with $K=0$ and $L^P=3^-$, $4^-$,
$5^-$, $\ldots$, with $K=3$. The three fold symmetry excludes states with $K=1$ and $K=2$
and leads to the lowest predicted $L^P = 4^\pm$ parity doublet in the $(v_1,0^{0})$
vibrational band. The predicted $L^P = 4^\pm$ parity doublet both in the ground band and the
Hoyle band is a strong signature of this model.
For the bending vibration with $(0,1^{1})$ the rotational
sequence is given by $L^{P}=1^{-}$, $2^-$, $3^-$, $4^-$, $\ldots$, with $K=1$,
$L^{P}=2^+$, $3^+$, $4^+$, $\ldots$, with $K=2$ and $L^{P}=4^{+}$, $\ldots$, with $K=4$.
The degeneracy of the states with the same value of the angular momentum $L$ but
different value of $K$ is split by the $\kappa_2$ term in Eq.~(\ref{ost}) \cite{Bi02}.
Since in the application to the cluster states of $^{12}$C, the vibrational and rotational
energies are of the same order, we expect sizeable rotation-vibration couplings.

\begin{figure}
\includegraphics[width=6in]{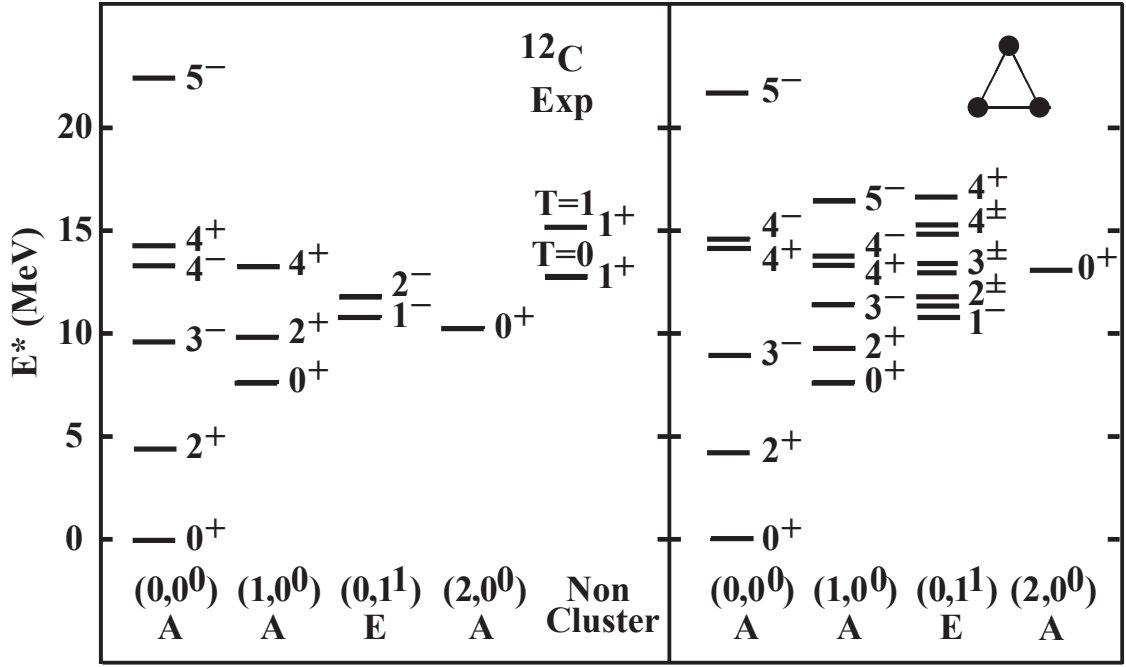}
\caption{\label{c12} Comparison between the low-lying experimental spectrum of $^{12}$C and the energies
of the oblate symmetric top calculated using Eq.~(\ref{ost}) with parameters that are
discussed in the text. The levels are organized in columns corresponding to the ground state band and the
vibrational bands with $A$ and $E$ symmetry of an oblate top with triangular symmetry. The last column on
the left-hand side, shows the lowest observed non-cluster ($1^+$) levels.}.

\end{figure}

In the $U(7)$ algebraic cluster model the energy eigenvalues of the oblate top, up to
terms quadratic in the rotation-vibration interaction, are given by:
\newpage
\ba
E = E_0 &+& \omega_{1}(v_{1}+\frac{1}{2}) \left( 1-\frac{v_1+1/2}{N} \right)
\nonumber\\
&+& \omega_{2}(v_{2}+1) \left( 1-\frac{v_2+1}{N+1/2} \right)
\nonumber\\
&+& \kappa_{1} \, L(L+1) + \kappa_{2} \, (K \mp 2\ell_{2})^{2}
\nonumber\\
&+& \left[\lambda_{1} \, (v_{1}+\frac{1}{2}) + \lambda_{2} \, (v_{2}+1) \right] L(L+1) ~.
\label{ost}
\ea
This formula includes both the anharmonicities which depend on $N$ and the vibrational
dependence of the moments of inertia. In Fig.~\ref{c12} we show a comparison of the
cluster states of $^{12}$C with the spectrum of the oblate top according to the approximate
energy formula of Eq.~(\ref{ost}). The coefficients $\kappa_1$, $\lambda_1$
and $\lambda_2$ are determined by the moments of inertia of the ground state band,
the Hoyle band and the bending vibration. The value of $\kappa_2$ term is determined
from the relative energies of the positive and negative parity states in the ground
state band. The vibrational energies $\omega_1$ and $\omega_2$ are obtained from the
excitation energies of the $0^+$ Hoyle state and the $1^-$ state, respectively.
Whereas in molecules the anharmonicities are small and hence $N$ is large, in $^{12}$C
the situation is completely different. The rotation-vibration couplings and anharmonicities
are large and therefore $N$ is small. Here it is taken to be $N=10$ \cite{Bi00,Bi02}.
The large anharmonicities lead to an increase of the rms radius of the vibrational
excitations relative to that of the ground state.

In this analysis the ground state rotational band of $^{12}$C is composed of both
positive parity with $J^\pi=0^+$, $2^+$, $4^+$ with $K=0$ and negative parity state
with $J^\pi = 3^-$, $4^-$, $5^-$ with $K=3$. Since these states belong to a single
rotational structure, the electromagnetic transition probabilities from the ground state
to the $2^+$ and $3^-$ (and $4^+$) states are related. The measured strong electromagnetic
transitions $B(E2;2^+ \rightarrow 0^+) = 4.9 \pm 0.3$ W.U. and
$B(E3;3^- \rightarrow 0^+) = 12.9 \pm 1.7$ W.U. indicate collectivity which is not predicted for simple shell model states. The agreement
with the predicted B(E$\lambda$) values for the $2^+$ and $3^-$ states of the ground state rotational band \cite{Bi02} and the agreement of the predicted form factors measured in $^{12}$C(e,e') scattering for $2^+, \ 3^-$ and the $4^+$ state \cite{Bi02} indicates that the $K=0$ and $K=3$ bands which are usually considered as separate bands, coalesce to form a single ground state rotational structure. As we
discussed above the merging of the $K=0$ and $K=3$ bands leads to the predicted $J^\pi = 4^\pm$
degenerate parity doublet which is a strong signature of the ${\cal D}_{3h}$ symmetry.

The $0^+$ Hoyle state in $^{12}$C at 7.654 MeV is interpreted as the band-head of the $A$ symmetric
stretching vibration or breathing mode of the triangular configuration with the same geometrical
arrangement and rotational structure as for the ground state rotational band, as shown in Fig.~\ref{c12}.
The non-harmonicity of the potential discussed above leads to larger rms radii for higher vibrational states,
hence the Hoyle rotational band is predicted to have a moment of inertia larger than the ground state band
(by a factor of 2). Recent measurements revealed  the existence of the $2^+$ \cite{2+} and $4^+$ \cite{Fre11}
members of the Hoyle rotational band which raises the question of the identification
of the predicted negative parity states shown in
Fig.~\ref{c12}. The $4^-$ state which is predicted to be nearly degenerate with the $4^+$ state,
can be measured for example in 180$^\circ$ electron scattering off $^{12}$C \cite{Darmst}.
We note that a (broad) $3^-$ state was suggested to lie between 11 and 14 MeV \cite{Fre07}
which is close to the predicted energy shown in Fig.~\ref{c12}. In order to distinguish
between different geometric configurations of the Hoyle band, {\it e.g.} equilateral triangular or bent-arm,
the identification of the negative parity states $3^-$ and $4^-$ is crucial which is a
strong motivation for a dedicated experimental search \cite{Darmst}.

The $1^-$ state at 10.84 MeV is assigned as the bandhead of the vibrational bending mode
whose lowest-lying rotational excitations consist of nearly degenerate parity doublets of
$2^\pm$ and $3^\pm$ states. So far, only the $2^-$ has been identified.

In addition to the cluster states, there are other (non cluster) states in $^{12}$C. In particular, with $3\alpha$ configurations no $1^+$ can be formed. The two $1^+$ ($T=0$) and $1^+$ ($T=1$) states at 12.71 MeV and 15.11 MeV, respectively, shown in Fig. 5, are therefore clearly non-cluster states and indicate the energy above which the identification of cluster or non-cluster low spin states becomes difficult.
The cluster states are characterized by large alpha widths ($\Gamma_{\alpha 0}$, the decay to the ground
state of $^8$Be or $\Gamma_{\alpha 1}$, the decay to the first excited $2^+$ state of $^8$Be) with
reduced widths that exhaust a large fraction of the Wigner limit.

Of particular interest \cite{Physics} is the Hoyle state with a geometrical arrangement of the
alpha-particles that may be deduced from the rotational band built on top of the Hoyle state
\cite{2+,Physics,Fre11}. While the observed moment of inertia of the Hoyle band excludes the proposed
linear chain structure of the Hoyle state \cite{2+} two geometrical alternatives of either equilateral
triangular arrangements or obtuse triangular arrangement are considered for the arrangement of the
three alpha particles in the Hoyle state of $^{12}$C and can be resolved by the future measurements we propose here.

In conclusion we presented evidence for triangular ${\cal D}_{3h}$ symmetry in the arrangement
of the three alpha-particle in the ground state of $^{12}$C. Such a symmetry is now established in molecular physics and 
nuclear physics. Another interesting application would be
to odd-mass nuclei. Finally, the algebraic cluster model predicts several additional state. The predicted
broad and overlapping states require accurate data as measured for the second $2^+$ Hoyle state \cite{2+}.
The selectivity of gamma-ray beams \cite{2+} as well as electron beams \cite{Darmst} would aid in populating the
states of interest and resolve the broad interfering states. These new capabilities should initiate an 
extensive experimental program for the search of the predicted
(``missing") states and promises to shed new light on the clustering  phenomena in light nuclei.

\vspace{-0.5cm}
\section*{Acknowledgement}

The authors wish to acknowledge extensive discussions with Professor Francesco Iachello and thank him for stimulating this study. This work is supported in part by the U.S. Department of Energy, Grant Number DE-FG02-94ER40870, and in part by research projects from DGAPA-UNAM and CONACyT, Mexico.

\end{document}